\documentstyle[12pt]{article}

\begin{document}

\title{\Large {\bf Commutators and propagators of Moyal star-products 
                   and microcausality for free scalar field on 
                   noncommutative spacetime } }
  
\author{{{\large Zheng Ze Ma}} \thanks{Electronic address: 
           z.z.ma@seu.edu.cn} 
  \\  \\ {\normalsize {\sl Department of Physics, Southeast University, 
         Nanjing, 210096, P. R. China } }}

\date{}

\maketitle

\begin{abstract}

\indent
 
  We study the Moyal commutators and their expectation values between 
vacuum states and non-vacuum states for free scalar field on 
noncommutative spacetime. Then from the Moyal commutators, we find 
that the microcausality is satisfied for the linear operators of the 
free scalar field on noncommutative spacetime. We construct the Feynman 
propagator of Moyal star-product for noncommutative scalar field theory.

\end{abstract}

~~~ PACS numbers: 11.10.Nx, 03.70.+k

\section{Introduction} 

\indent

  Spacetime may have discrete and noncommutative structures under a 
very small microscopic scale. The concept of spacetime noncommutativity 
was first proposed by Snyder [1]. About ten years ago, Doplicher 
{\sl et al}. proposed the uncertainty relations for the measurement of 
spacetime coordinates and the simplified noncommutative algebra for the 
spacetime coordinates [2]. In recent years, spacetime noncommutativity 
was discovered again in superstring theories [3]. The end points of 
open strings trapped on a $D$-brane with a nonzero NSNS two form 
background $B$-field turn out to be noncommutative, noncommutative 
field theories then appear as the low energy effective theory of these 
$D$-branes. It has resulted a lot of researches on noncommutative field 
theories [4,5].

  In noncommutative spacetime, we can regard the Moyal star-product as 
the basic product operation. Thus it is necessary to study the 
commutators and propagators of Moyal star-products for quantum fields 
on noncommutative spacetime. In this paper we will study the commutators 
and propagators of Moyal star-products for scalar field on noncommutative 
spacetime. In Sec. II, we construct the Moyal commutators and evaluate 
their expectation values between vacuum states and non-vacuum states for 
noncommutative scalar field theory. We find that the microcausality is 
satisfied for the linear operators of the free scalar field on 
noncommutative spacetime. In Sec. III, we calculate the Feynman 
propagator of Moyal star-product for scalar field on noncommutative 
spacetime. In Sec. IV, we discuss some of the problems for the 
microcausality of noncommutative scalar field theory.

\section{ Commutators of Moyal star-products and their 
          expectation values }

\indent
 
  In noncommutative spacetime, spacetime coordinates satisfy the 
commutation relation
$$
  [x^{\mu},x^{\nu}]=i\theta^{\mu\nu} ~,
  \eqno{(2.1)}  $$
where $\theta^{\mu\nu}$ is a constant real antisymmetric matrix that 
parameterizes the noncommutativity of the spacetime. For field theories 
on noncommutative spacetime, they can be obtained through introducing the 
Moyal star-product, i.e., all of the products between field functions are 
replaced by the Moyal star-products in the Lagrangian. The Moyal 
star-product of two functions $f(x)$ and $g(x)$ is defined to be 
\begin{eqnarray*}
  f(x)\star g(x) & = & e^{\frac{i}{2}\theta^{\mu\nu}\frac{\partial}{\partial 
                 \alpha^\mu}\frac{\partial}{\partial\beta^{\nu}}}f(x+\alpha)
                 g(x+\beta)\vert_{\alpha=\beta=0} \\
                 & = & f(x)g(x)+\sum\limits^{\infty}_{n=1}
                 \left(\frac{i}{2}\right)^{n}
                 \frac{1}{n!}\theta^{\mu_{1}\nu_{1}}
                 \cdots\theta^{\mu_{n}\nu_{n}}
                 \partial_{\mu_{1}}\cdots\partial_{\mu_{n}}f(x) 
                 \partial_{\nu_{1}}\cdots\partial_{\nu_{n}}g(x) ~.
\end{eqnarray*}
$$ 
  \eqno{(2.2)}  $$
The Moyal star-product of two functions of Eq. (2.2) is defined at the 
same spacetime point. We can generalize Eq. (2.2) to two functions on 
different spacetime points [5]:
\begin{eqnarray*}
  f(x_{1})\star g(x_{2}) & = & e^{\frac{i}{2}
          \theta^{\mu\nu}\frac{\partial}{\partial 
          \alpha^\mu}\frac{\partial}{\partial\beta^{\nu}}}f(x_{1}+\alpha)
                 g(x_{2}+\beta)\vert_{\alpha=\beta=0} \\
                 & = & f(x_{1})g(x_{2})+\sum\limits^{\infty}_{n=1}
                 \left(\frac{i}{2}\right)^{n}
                 \frac{1}{n!}\theta^{\mu_{1}\nu_{1}}
                 \cdots\theta^{\mu_{n}\nu_{n}}
                 \partial_{\mu_{1}}\cdots\partial_{\mu_{n}}f(x_{1}) 
                 \partial_{\nu_{1}}\cdots\partial_{\nu_{n}}g(x_{2}) ~. 
\end{eqnarray*}
$$ 
  \eqno{(2.3)}  $$
Equation (2.3) can be established through generalizing the commutation 
relation of spacetime coorinates at the same point to two different 
points: 
$$
  [x_{1}^{\mu},x_{2}^{\nu}]=i\theta^{\mu\nu} ~. 
  \eqno{(2.4)}  $$
We can also expect to search for a grounds of argument for Eq. (2.4) 
from superstring theories.

  In noncommutative spacetime, we can regard the Moyal star-product 
as the basic product operation. Thus it is necessary to analyze the 
the commutators of Moyal star-products for quantum fields on 
noncommutative spacetime. To consider the noncommutative 
$\varphi^{\star4}$ scalar field theory, its Lagrangian is given by 
$$
  {\cal L}=\frac{1}{2}\partial^{\mu}\varphi\star\partial_{\mu}\varphi-
           \frac{1}{2}m^{2}\varphi\star\varphi-\frac{1}{4!}\lambda
             \varphi\star\varphi\star\varphi\star\varphi ~.
  \eqno{(2.5)}  $$ 
The Fourier expansion of the free scalar field is given by 
$$
  \varphi({\bf x},t)=\int\frac{d^{3}k}{\sqrt{(2\pi)^{3}2\omega_{k}}}
       [a(k)e^{i{\bf k}\cdot{\bf x}-i\omega t}+
        a^{\dagger}(k)e^{-i{\bf k}\cdot{\bf x}+i\omega t}] ~.
  \eqno{(2.6)}  $$
Here we adopt the usual Lorentz invariant spectral measure for the 
Fourier expansion of the scalar field [6,7]. In Eq. (2.6), we take the 
spacetime coordinates to be noncommutative. They satisfy the commutation 
relations (2.1) and (2.4). The commutation relations for the creation 
and annihilation operators are still the same as that in the ordinary 
commutative spacetime: 
$$
  [a(k),a^{\dagger}(k^{\prime})]=
        \delta^{3}({\bf k}-{\bf k}^{\prime}) ~,          $$
$$ 
  [a(k),a(k)]=0 ~,      $$
$$
  [a^{\dagger}(k),a^{\dagger}(k)]=0 ~.              
  \eqno{(2.7)}  $$
We define the commutator of the Moyal star-product to be 
$$
  [\varphi(x),\varphi(y)]_{\star}=\varphi(x)\star\varphi(y)-
              \varphi(y)\star\varphi(x) ~.
  \eqno{(2.8)}  $$
We call Eq. (2.8) the Moyal commutator for convenience. From the Fourier 
expansion of Eq. (2.6) for the free field, we can calculate the Moyal 
commutator of two scalar fields. It is given by 
\begin{eqnarray*}
  [\varphi(x),\varphi(y)]_{\star}
           & = & \int\frac{d^{3}kd^{3}k^{\prime}}
                {(2\pi)^{3}\sqrt{2\omega_{k}2\omega_{k}^{\prime}}} 
                \left[a(k)e^{-ikx}+a^{\dagger}(k)e^{ikx},  
                 a(k^{\prime})e^{-ik^{\prime}y}+a^{\dagger}
                 (k^{\prime})e^{ik^{\prime}y}\right]_\star      \\ 
           & = & \int\frac{d^{3}kd^{3}k^{\prime}}
         {(2\pi)^{3}\sqrt{2\omega_{k}2\omega_{k}^{\prime}}}\Bigg\{
      \left[a(k)e^{-ikx},a(k^{\prime})e^{-ik^{\prime}y}\right]_\star+
      \left[a(k)e^{-ikx},a^{\dagger}(k^{\prime})e^{ik^{\prime}y}
               \right]_\star               \\
           & ~ & ~~~~~~~~~ +\left[a^{\dagger}(k)e^{ikx},a(k^{\prime})
                      e^{-ik^{\prime}y}\right]_\star+
                 \left[a^{\dagger}(k)e^{ikx},a^{\dagger}
                 (k^{\prime})e^{ik^{\prime}y}\right]_\star\Bigg\} ~.
\end{eqnarray*}      
$$  \eqno{(2.9)}  $$
For the reason that there are two kinds of noncommutative structures,  
i.e., field operators and spacetime coordinates, the spacetime coordinates 
now is noncommutative, we cannot apply the commutation relations for the 
creation and annihilation operators of Eq. (2.7) directly to obtain a 
$c$-number result for the Moyal commutator.

  In order to obtain the $c$-number result for the Moyal commutator, we can 
calculate its vacuum state expectation value. We have 
\begin{eqnarray*}
   & ~ & \langle0\vert[\varphi(x),\varphi(y)]
            _{\star}\vert0\rangle              \\
           & = & \langle0\vert\int\frac{d^{3}kd^{3}k^{\prime}}
            {(2\pi)^{3}\sqrt{2\omega_{k}2\omega_{k}^{\prime}}}   
        \left(a(k)a^{\dagger}(k^{\prime})e^{-ikx}\star e^{ik^{\prime}y}   
         -a(k^{\prime})a^{\dagger}(k)e^{-ik^{\prime}y}\star e^{ikx}
          \right)\vert0\rangle   \\
           & = & \int\frac{d^{3}k}{(2\pi)^{3}2\omega_{k}}   
                 \left[e^{-ikx}\star e^{iky}-
                  e^{-iky}\star e^{ikx}\right]       \\ 
           & = & \int\frac{d^{3}k}{(2\pi)^{3}2\omega_{k}}   
                 \left[\exp(\frac{i}{2}k\times k)e^{-ik(x-y)}
                   -\exp(\frac{i}{2}k\times k)e^{ik(x-y)}\right] ~,  
\end{eqnarray*}        
$$  \eqno{(2.10)}  $$
where we have applied Eq. (2.3) and we note 
$$
  k\times p=k_{\mu}\theta^{\mu\nu}p_{\nu} ~.
  \eqno{(2.11)}  $$
Because $\theta^{\mu\nu}$ is antisymmetric, $k\times k=0$, we obtain
\begin{eqnarray*}
  & ~ & \langle0\vert[\varphi(x),\varphi(y)]_{\star}\vert0\rangle
        =\int\frac{d^{3}k}{(2\pi)^{3}2\omega_{k}}   
                 \left[e^{-ik(x-y)}-e^{ik(x-y)}\right]            \\
  & = & -\frac{i}{(2\pi)^{3}}\int\frac
                    {d^{3}k}{\omega_{k}}e^{i{\bf k}\cdot({\bf x}-
       {\bf y})}\sin\omega_{k}(x_{0}-y_{0}) = i\Delta(x-y) ~. 
\end{eqnarray*}      
$$ 
  \eqno{(2.12)}  $$
So the result of Eq. (2.12) is equal to the commutator of scalar field 
in ordinary commutative spacetime: 
$$
  [\varphi(x),\varphi(y)]=-\frac{i}{(2\pi)^{3}}\int\frac
           {d^{3}k}{\omega_{k}}e^{i{\bf k}\cdot({\bf x}-{\bf y})}
           \sin\omega_{k}(x_{0}-y_{0}) = i\Delta(x-y) ~.      
  \eqno{(2.13)}  $$ 
It is obvious to see that this equality relies on the antisymmetry of 
$\theta^{\mu\nu}$.

  We can also calculate the expectation values between non-vacuum states 
for the Moyal commutators. Let $\vert\Psi\rangle$ represent a normalized 
non-vacuum physical state which is in the occupation eigenstate:
$$
  \vert\Psi\rangle=\vert N_{k_{1}}N_{k_{2}}\cdots N_{k_{i}}\cdots,0
              \rangle ~,
  \eqno{(2.14)}  $$ 
where $N_{k_{i}}$ represents the occupation number of the momentum $k_{i}$. 
We can suppose that the occupation numbers are nonzero only on some 
separate momentums $k_{i}$. For all other momentums, the occupation 
numbers are zero. We use $0$ to represent that the occupation numbers are 
zero on all the other momentums in Eq. (2.14). The state vector 
$\vert\Psi\rangle$ has the following properties: 
$$  
  \langle N_{k_{1}}N_{k_{2}}\cdots N_{k_{i}}\cdots\vert
          N_{k_{1}}N_{k_{2}}\cdots N_{k_{i}}\cdots\rangle =1 ~,     
  \eqno{(2.15)}  $$   
$$
  \sum\limits_{N_{k_{1}}N_{k_{2}}\cdots}
    \vert N_{k_{1}}N_{k_{2}}\cdots N_{k_{i}}\cdots\rangle
    \langle N_{k_{1}}N_{k_{2}}\cdots N_{k_{i}}\cdots\vert=1 ~,
  \eqno{(2.16)}  $$   
$$
  a(k_{i})\vert N_{k_{1}}N_{k_{2}}\cdots N_{k_{i}}\cdots\rangle=
          \sqrt{N_{k_{i}}}\vert N_{k_{1}}N_{k_{2}}\cdots 
                (N_{k_{i}}-1)\cdots\rangle ~,           
  \eqno{(2.17)}  $$ 
$$
  a^{\dagger}(k_{i})\vert N_{k_{1}}N_{k_{2}}\cdots N_{k_{i}}
          \cdots\rangle=\sqrt{N_{k_{i}}+1}\vert N_{k_{1}}N_{k_{2}}
          \cdots(N_{k_{i}}+1)\cdots\rangle ~.
  \eqno{(2.18)}  $$ 
Equation (2.16) is the completeness expression for the state vector 
$\vert\Psi\rangle$. Therefore Eq. (2.14) can represent an arbitrary 
scalar field quantum sysytem. Then from Eq. (2.9), the expectation 
value between any non-vacuum physical states for the Moyal commutator 
is given by   
\begin{eqnarray*}
  & ~ & \langle\Psi\vert[\varphi(x),\varphi(y)]_{\star}
               \vert\Psi\rangle              \\
  & = & \langle\Psi\vert\int\frac{d^{3}kd^{3}k^{\prime}}
         {(2\pi)^{3}\sqrt{2\omega_{k}2\omega_{k}^{\prime}}}\left\{
          \left[a(k)e^{-ikx},a^{\dagger}(k^{\prime})e^{ik^{\prime}y}
               \right]_\star              
        +\left[a^{\dagger}(k)e^{ikx},a(k^{\prime})
         e^{-ik^{\prime}y}\right]_\star\right\}\vert\Psi\rangle   \\
  & = & \int\frac{d^{3}k}{(2\pi)^{3}2\omega_{k}}   
                 \left[e^{-ikx}\star e^{iky}-
                 e^{-iky}\star e^{ikx} \right]  
            =i\Delta(x-y) ~,
\end{eqnarray*}  
$$  \eqno{(2.19)}  $$
which is just equal to the vacuum state expectation value of Eq. (2.12). 
The properties of the commutation relations for the creation and 
annihilation operators of Eq. (2.7) are still reflected in the above 
evaluations for the non-vacuum state expectation values for the Moyal 
commutators.

  The vacuum state expectation value and non-vacuum state expectation 
value of the equal-time Moyal commutator can be obtained from 
Eqs. (2.12) and (2.19). They are 
$$
  \langle0\vert[\varphi({\bf x},t),\varphi({\bf y},t)]_{\star}
          \vert0\rangle=\Delta({\bf x}-{\bf y},0)=0 ~,       
  \eqno{(2.20)}  $$ 
$$
  \langle\Psi\vert[\varphi({\bf x},t),\varphi({\bf y},t)]_{\star}
          \vert\Psi\rangle=\Delta({\bf x}-{\bf y},0)=0 ~.      
  \eqno{(2.21)}  $$ 
Because $\Delta(x-y)$ is a Lorentz invariant singular function, 
it satisfies 
$$
  \Delta(x-y)=0 ~~~~~~ \mbox{for all} ~~~~~~ (x-y)^{2}<0 ~,
  \eqno{(2.22)}  $$
we have 
$$
  \langle0\vert[\varphi(x),\varphi(y)]_{\star}\vert0\rangle=0
           ~~~~~~ \mbox{for all} ~~~~~~ (x-y)^{2}<0 ~, 
  \eqno{(2.23)}  $$
and similarly 
$$
  \langle\Psi\vert[\varphi(x),\varphi(y)]_{\star}
             \vert\Psi\rangle=0
           ~~~~~~ \mbox{for all} ~~~~~~ (x-y)^{2}<0 ~. 
  \eqno{(2.24)}  $$
For quantum fields in ordinary commutative spacetime, they 
satisfy the microcausality principle [6,7]. For scalar field theory 
we have  
$$
  [\varphi(x),\varphi(y)]=0
           ~~~~~~ \mbox{for all} ~~~~~~ (x-y)^{2}<0 ~,
  \eqno{(2.25)}  $$
which means that any two fields as physical observables commute with 
each other when they are separated by a spacelike interval. Or we can 
say any two physical measurements separated by a spacelike 
interval do not interfere each other. In noncommutative spacetime, 
we can regard the Moyal star-product as the basic product operation. 
Although we cannot obtain that 
$[\varphi(x),\varphi(y)]_{\star}=0$ for $(x-y)^{2}<0$ from Eqs. (2.23) 
and (2.24), because any physical measurement is taken under certain 
physical state, what the observer measures are certain expectation 
values in fact, Eqs. (2.23) and (2.24) can still represent the 
satisfying of microcausality for free scalar field on 
noncommutative spacetime. However in the above we have only analyzed 
the linear operator $\varphi(x)$. For the other linear operators of 
the scalar field such as $\partial^{\mu}\varphi(x)$, we can obtain that 
their microcausality properties on noncommutative spacetime are similar 
to that of the linear operator $\varphi(x)$. They also satisfy the 
microcausality. The calculation and result are similar as above. Thus 
we omit to write down them here explicitly. For the quadratic operators 
of free scalar field on noncommutative spacetime such as 
$\varphi(x)\star\varphi(x)$, their microcausality properties need to 
be studied further.

\section{ Feynman propagator of Moyal star-product 
                      for noncommutative scalar field theory } 

\indent

  The same as quantum fields in ordinary commutative spacetime [6,7], 
we can decompose the Fourier expansion of the free scalar field 
into positive frequency part and negative frequency part: 
$$
  \varphi(x)=\varphi^{+}(x)+\varphi^{-}(x) ~,              
  \eqno{(3.1)}  $$ 
where 
$$
  \varphi^{+}(x)=\int\frac{d^{3}k}{\sqrt{(2\pi)^{3}2\omega_{k}}}
           a(k)e^{i{\bf k}\cdot{\bf x}-i\omega t}                         
         =\int\frac{d^{3}k}{\sqrt{(2\pi)^{3}2\omega_{k}}}
           a(k)e^{-ikx} ~,                   
  \eqno{(3.2)}  $$ 
$$          
  \varphi^{-}(x)=\int\frac{d^{3}k}{\sqrt{(2\pi)^{3}2\omega_{k}}}
             a^{\dagger}(k)e^{-i{\bf k}\cdot{\bf x}+i\omega t}
            =\int\frac{d^{3}k}{\sqrt{(2\pi)^{3}2\omega_{k}}}
             a^{\dagger}(k)e^{ikx} ~.                 
  \eqno{(3.3)}  $$ 
According to Eq. (2.9), we can decompose the vacuum expectation value 
of the Moyal commutator into two parts: 
$$
  \langle0\vert[\varphi(x),\varphi(y)]_{\star}\vert0\rangle
        =\langle0\vert[\varphi^{+}(x),\varphi^{-}(y)]_{\star}
            \vert0\rangle+
         \langle0\vert[\varphi^{-}(x),\varphi^{+}(y)]_{\star}
            \vert0\rangle ~.       
  \eqno{(3.4)}  $$ 
For the two parts of Eq. (3.4), we obtain 
\begin{eqnarray*}
  & ~ & \langle0\vert[\varphi^{+}(x),\varphi^{-}(y)]_{\star}
        \vert0\rangle = \langle0\vert
        \int\frac{d^{3}k}{(2\pi)^{3}2\omega_{k}}
         [a(k)e^{-ikx},a^{\dagger}(k)e^{iky}]
         _{\star}\vert0\rangle            \\
           & = & \int\frac{d^{3}k}{(2\pi)^{3}2\omega_{k}}   
                 e^{-ikx}\star e^{iky}=
         \int\frac{d^{3}k}{(2\pi)^{3}2\omega_{k}}e^{-ik(x-y)}   
             = i\Delta^{+}(x-y) ~,
\end{eqnarray*} 
$$  \eqno{(3.5)}  $$ 
and 
\begin{eqnarray*}
  & ~ & \langle0\vert[\varphi^{-}(x),\varphi^{+}(y)]_{\star}
        \vert0\rangle = \langle0\vert
        \int\frac{d^{3}k}{(2\pi)^{3}2\omega_{k}}
         [a^{\dagger}(k)e^{ikx},a(k)e^{-iky}]
         _{\star}\vert0\rangle            \\
           & = & -\int\frac{d^{3}k}{(2\pi)^{3}2\omega_{k}}   
                 e^{-iky}\star e^{ikx}=
         -\int\frac{d^{3}k}{(2\pi)^{3}2\omega_{k}}e^{-ik(y-x)}  \\ 
           & = & -i\Delta^{+}(y-x)=i\Delta^{-}(x-y) ~.
\end{eqnarray*} 
$$  \eqno{(3.6)}  $$ 
Thus we can write
$$
  \langle0\vert[\varphi(x),\varphi(y)]_{\star}\vert0\rangle
      =i\Delta(x-y)=i(\Delta^{+}(x-y)+\Delta^{-}(x-y)) ~.             
  \eqno{(3.7)}  $$ 
We can also see that if we replace the vacuum state by any 
non-vacuum state in the above formulas the results will not change. 
The above results are also equal to the corresponding cummutators 
in the ordinary commutative spacetime.

  For the commutator (3.5), we can rewrite it furthermore:
\begin{eqnarray*}
  & ~ & \langle0\vert[\varphi^{+}(x),\varphi^{-}(y)]_{\star}
        \vert0\rangle = \langle0\vert\varphi^{+}(x)\star
        \varphi^{-}(y)\vert0\rangle       \\ 
  & = & \langle0\vert\varphi(x)\star\varphi(y)\vert0\rangle
        = i\Delta^{+}(x-y) ~. 
\end{eqnarray*} 
$$  \eqno{(3.8)}  $$ 
We define the time-ordered Moyal star-product of two scalar 
field operators to be 
$$
  T\varphi(x)\star\varphi(x^{\prime})=\theta(t-t^{\prime})
         \varphi(x)\star\varphi(x^{\prime})+\theta(t^{\prime}-t)
         \varphi(x^{\prime})\star\varphi(x) ~,                    
  \eqno{(3.9)}  $$
where $\theta(t-t^{\prime})$ is the unit step function. We can 
calculate the vacuum expectation value of Eq. (3.9): 
$$
  \langle0\vert T\varphi(x)\star\varphi(x^{\prime})\vert0\rangle
         =\theta(t-t^{\prime})\langle0\vert\varphi(x)\star
          \varphi(x^{\prime})\vert0\rangle+\theta(t^{\prime}-t)
          \langle0\vert\varphi(x^{\prime})
          \star\varphi(x)\vert0\rangle ~.                  
  \eqno{(3.10)}  $$
Equation (3.10) is just the Feynman propagator of Moyal star-product 
for scalar field. We can call it the Feynman Moyal propagator for 
convenience. To introduce the singular function $\Delta_{F}(x)$, 
we can write the Feynman Moyal propagator (3.10) as  
$$
  \langle0\vert T\varphi(x)\star\varphi(x^{\prime})
        \vert0\rangle=i\Delta_{F}(x-x^{\prime}) ~.     
  \eqno{(3.11)}  $$
From Eqs. (3.6), (3.8), and (3.10), we have 
$$
  \Delta_{F}(x-x^{\prime})=\theta(t-t^{\prime})
       \Delta^{+}(x-x^{\prime})-\theta(t^{\prime}-t)
       \Delta^{-}(x-x^{\prime}) ~,                        
  \eqno{(3.12)}  $$
where the momentum integral representation of the singular function 
$\Delta_{F}(x-x^{\prime})$ is given by 
$$
  \Delta_{F}(x-x^{\prime})=\int\frac{d^{4}k}{(2\pi)^{4}}
       \frac{1}{k^{2}-m^{2}+i\epsilon}e^{-ik\cdot(x-x^{\prime})} ~.  
  \eqno{(3.13)}  $$
From the above results, we can see the Feynman Moyal propagator of 
noncommutative scalar field is just equal to the Feynman propagator 
of scalar field on ordinary commutative spacetime. However it is 
necessary to point out that in Eq. (3.9) for the definition of the 
time-ordered Moyal star-product of two scalar fields, we have applied 
a simplified manipulation. This is because the Moyal star-products 
are not invariant generally under the exchange of the orders of two 
functions, for the second term of the right hand side of Eq. (3.9) we 
need to consider this fact. In the Fourier integral representation, 
it can be seen clearly that the second term of the right hand side of 
Eq. (3.9) will have an additional phase factor $e^{ik\times k^{\prime}}$ 
in contrast to the first term of the right hand side of Eq. (3.9) due 
to the exchange of the order of $\varphi(x)$ and $\varphi(x^{\prime})$ 
for the Moyal star-product. However in Eq. (3.10) when we calculate 
the vacuum expectation value for Eq. (3.9), we can see that the 
non-zero contribution comes from the $k=k^{\prime}$ part inside the 
integral (cf. Eqs. (3.5) and (3.6)). This will make the phase factor 
to be $e^{ik\times k}$, which is $1$ due to the antisymmetry of 
$\theta^{\mu\nu}$. Thus in the right hand side of Eq. (3.9), we can 
omit this effect in the exchange of the order of two scalar fields 
for their Moyal star-product equivalently.

  Just like that in ordinary commutative spacetime, the physical 
meaning of the Feynman Moyal propagator (3.10) can also be explained 
as the vacuum to vacuum transition amplitude for quantum fields on 
noncommutative spacetime. We can also construct the Feynman Moyal 
propagators for other fields such as the fermion field and 
electromagnetic field. The reason why we would like to construct 
the Feynman propagators of Moyal star-products for quantum fields on 
noncommutative spacetime is that: for noncommutative field theories 
we can establish their $S$-matrix where the products between field 
operators in ${\cal H}_{int}$ are Moyal star-products. From the 
Wick's theorem expansion for the time ordered products of field 
operators, there will occur the Feynman Moyal propagators. Therefore 
it is necessary to study Feynman propagators of Moyal star-products 
for quantum fields on noncommutative spacetime.

\section{ Discussion }

\indent

  In this paper, we studied the Moyal commutators and their 
expectation values between vacuum states and non-vacuum states 
for free scalar field on noncommutative spacetime. We also 
studied the Feynman propagator of Moyal star-product for scalar 
field on noncommutative spacetime. From the Moyal commutators, 
we find that the microcausality is satisfied on noncommutative 
spacetime for the linear operators of the free scalar field. 
For the quadratic operators of the free scalar field on 
noncommutative spacetime, their microcausality properties need to 
be studied further. For the microcausality of the quadratic 
operators of the scalar field on noncommutative spacetime, their 
microcausality properties have been studied by some authors in the 
literature [8,9]. In Ref. [8], Greenberg obtained that for the 
commutators 
$[:\varphi(x)\star\varphi(x):,:\varphi(y)\star\varphi(y):]_{\star}$ 
and $[:\varphi(x)\star\varphi(x):,\partial_{\mu}:\varphi(y)
\star\varphi(y):]$, they do not satisfy the microcausality on 
noncommutative spacetime. However the result of Ref. [8] is based on 
an incomplete expansion and quantization of the scalar field. 
In Ref. [9], through analyzing the expectation value 
$\langle0\vert[:\varphi(x)\star\varphi(x):,:\varphi(y)\star\varphi(y):]
\vert p,p^{\prime}\rangle$, Chaichian {\sl et al}. obtained that 
microcausality is violated for quantum fields on noncommutative 
spacetime when $\theta^{0i}\neq0$. However, for the microcausality 
problem, it is more reasonable that we should analyze the expectation 
values between the same state vectors for the field operators. 
In Ref. [10], through supposing the spectral measure to be the form 
of $SO(1,1)\times SO(2)$ invariance, the authors obtained the result 
that microcausality is violated for quantum fields on noncommutative 
spacetime generally. However, such a conclusion is a necessary result 
of the breakdown of the Lorentz invariance in Ref. [10]. In Ref. [11], 
the authors obtained that even the $SO(1,1)$ microcausality may be 
violated for quantum fields on noncommutative spacetime through 
calculating the propagators with quantum corrections.

  It is necessary to point out that the microcausality problem 
discussed in this paper is only restricted to free fields. For quantum 
field theories on noncommutative spacetime, there exist some other 
possibilities to violate the microcausality. For example because there 
exist the UV/IR mixing phenomena [12], the infrared singularities that 
come from non-planar diagrams may result the existence of nonlocal and 
instantaneous interactions [13]. These nonlocal and instantaneous 
interactions do not satisfy the usual microcausality of quantum field 
theories on ordinary commutative spacetime. For the classical theory of 
fields on noncommutative spacetime, there also exist the possibilities 
that violate the causality principle. For example in Ref. [14], Durhuus 
and Jonsson proved that for nonlinear noncommutative waves, their 
propagation speed is infinite. They found that if the initial conditions 
have a compact support then for any positive time the support of the 
solution can be arbitrarily large. In Ref. [15], the authors have found 
that for the solitons in noncommutative gauge theories, they can travel 
faster than the speed of light with an arbitrary speed. These phenomena 
mean that for noncommutative field theories, there exist the 
instantaneous interactions even though for their classical 
field theories.

\vskip 1cm

\noindent {\large {\bf References}}

\vskip 12pt

[1] H.S. Snyder, Phys. Rev. {\bf 71}, 38 (1947). 

[2] S. Doplicher, K. Fredenhagen, and J.E. Roberts, Phys. Lett. B 
    {\bf 331}, 39 (1994); 

    ~~~ Commun. Math. Phys. {\bf 172}, 187 (1995), hep-th/0303037. 
    
[3] N. Seiberg and E. Witten, J. High Energy Phys. 09 (1999) 032, 
    hep-th/9908142,  

    ~~~ and references therein. 
 
[4] M.R. Douglas and N.A. Nekrasov, Rev. Mod. Phys. {\bf 73}, 977 
    (2001), hep-th/0106048. 

[5] R.J. Szabo, Phys. Rep. {\bf 378}, 207 (2003), hep-th/0109162. 

[6] J.D. Bjorken and S.D. Drell, {\sl Relativistic Quantum Fields}  
    (McGraw-Hill, 1965). 

[7] C. Itzykson and J.-B. Zuber, {\sl Quantum Field Theory} 
    (McGraw-Hill Inc., 1980). 

[8] O.W. Greenberg, hep-th/0508057. 

[9] M. Chaichian, K. Nishijima, and A. Tureanu, Phys. Lett. B 
    {\bf 568}, 146 (2003), hep-

    ~~~ th/0209008. 

[10] L. {\' A}lvarez-Gaum{\' e} and M.A. V{\' a}zquez-Mozo, Nucl. Phys.
     {\bf B668}, 293 (2003), hep-

     ~~~~~ th/0305093. 

[11] L. Alvarez-Gaum{\' e}, J.L.F. Barb{\' o}n, and R. Zwicky, 
     J. High Energy Phys. 05 (2001)

     ~~~~~ 057, hep-th/0103069.  

[12] S. Minwalla, M. Van Raamsdonk, and N. Seiberg, J. High Energy 
     Phys. 02 (2000)

     ~~~~~ 020, hep-th/9912072.

[13] M. Van Raamsdonk, J. High Energy Phys. 11 (2001) 006, 
     hep-th/0110093.  

[14] B. Durhuus and T. Jonsson, J. High Energy Phys. 10 (2004) 050, 
     hep-th/0408190. 

[15] A. Hashimoto and N. Itzhaki, Phys.Rev. D {\bf 63}, 126004 (2001), 
     hep-th/0012093.

\end{document}